\documentclass[pra, twocolumn]{revtex4}
\usepackage[ansinew]{inputenc}
\usepackage{graphicx}
\usepackage{amsmath}
\usepackage{amsthm}
\usepackage{bm}
\usepackage{layout}
\usepackage{float}
\usepackage{amsfonts}
\usepackage{amssymb}
\usepackage{txfonts}%
\newcommand\id{\leavevmode\hbox{\small1\kern-3.3pt\normalsize1}}

\newcommand{\bra}{\langle}
\newcommand{\ket}{\rangle}

\newcommand{\tr}{\mbox{Tr}}

\begin{document}

\title{Operational quantum theory without predefined time
}

\author{Ognyan Oreshkov and Nicolas J. Cerf}

\affiliation{QuIC, Ecole Polytechnique de Bruxelles, CP 165, Universit\'{e} Libre de Bruxelles, 1050 Brussels, Belgium.}

\begin{abstract}

The standard formulation of quantum theory assumes a predefined notion of time. This is a major obstacle in the search for a quantum theory of gravity, where the causal structure of space-time is expected to be dynamical and fundamentally probabilistic in character. Here, we propose a generalized formulation of quantum theory without predefined time or causal structure, building upon a recently introduced operationally time-symmetric approach to quantum theory. The key idea is a novel isomorphism between transformations and states which depends on the symmetry transformation of time reversal. This allows us to express the time-symmetric formulation in a time-neutral form with a clear physical interpretation, and ultimately drop the assumption of time. In the resultant generalized formulation, operations are associated with regions that can be connected in networks with no directionality assumed for the connections, generalizing the standard circuit framework and the process matrix framework for operations without global causal order. The possible events in a given region are described by positive semidefinite operators on a Hilbert space at the boundary, while the connections between regions are described by entangled states that encode a nontrivial symmetry and could be tested in principle. We discuss how the causal structure of space-time could be understood as emergent from properties of the operators on the boundaries of compact space-time regions. The framework is compatible with indefinite causal order, timelike loops, and other acausal structures. 

\end{abstract}

\maketitle

\section*{Introduction}

\renewcommand{\figurename}{{Fig.}}
\renewcommand{\thefigure}{{\arabic{figure}}}

Quantum theory can be understood as a theory that prescribes probabilities for the outcomes of operations composed in different configurations. This perspective can be made precise in the framework of generalized probabilistic theories \cite{Ludwig, Hardy, Barrett, HardyCircuit,CDP2, DB, MM, CDP, Hardy2, infounit}, which provides the operational foundations of concepts such as states, effects, and transformations. In its standard form, however, this approach presupposes a definite causal structure and the notion of operation that it is based on assumes a notion of time direction. But where does this causal structure come from and what defines time’s direction? In the classical theory of general relativity, the causal structure of space-time is a dynamical variable that depends on the distribution of matter and energy. In a theory of quantum gravity, the causal structure is also expected to be dynamical and most generally allowed to exist in ‘superpositions’ of different alternatives \cite{hardyqg}. Is it possible to formulate an operational paradigm for quantum theory without prior concepts of time and causal structure, which allows such more general possibilities and within which we could understand the causal structure in our familiar regimes from more primitive concepts? 

Here, we propose an operational formulation of finite-dimensional quantum theory without any predefined time or causal structure, building upon a recently introduced operationally time-symmetric approach to quantum theory \cite{OC2}. A key observation behind this approach is that the notion of operation in the standard formulation depends on a predefined time both through the explicit assumption that operations have input and output systems, as well as through the implicit assumption that the implementation of an operation does not involve post-selection. More specifically, what we regard as an operation in the standard approach is a set of possible events between an input and an output system conditional only on information available before the time of the input---a property that arguably underlies the interpretation that an operation is something that we are able to `choose', unlike the outcome of an operation.  To remedy this asymmetry, a more general notion of operation was proposed in Ref.~\cite{OC2}, which permits realizations via both pre- or post-selection. In this approach, an operation is not assumed to be up to the ‘free choice’ of an experimenter, but simply describes knowledge about the possible events between an input and an output system, conditional on local information. In the present paper, we extend this idea to arbitrary regions, developing a new formalism that allows us to express the theory without reference to any prior notion of time. 

Our key insight is a novel isomorphism between effects and transformations, similar in spirit to the Choi-Jamio{\l}kowski isomorphism \cite{jam, choi}, but crucially dependent on the form of time reversal, which provides it with a physical interpretation. Using this isomorphism, we first recast the time-symmetric circuit formulation of quantum theory in a time-neutral form. In this representation, each wire in a circuit is represented by a pair of systems instead of a single system, where the two systems support an entangled state. This state encodes the symmetry transformation of time reversal and could be measured in principle. Operations are described by collections of positive semidefinite operators, which are contracted with the entangled states to yield numbers that enter in the calculation of probabilities according to a generalization of Born's rule.  

Using this time-neutral formalism, we extend the circuit formulation of quantum theory to circuits that can contain cycles. This also gives rise to an extension of the process matrix formalism \cite{OCB}, which comes with an intuitive interpretation: every operation is seen as a destructive measurement on two input systems---one from the past and one from the future---while the generalized process matrix describes a quantum state on which the local measurements are applied. We show that the circuit framework permitting cycles and the extended process matrix framework are not only mathematically equivalent, but also operationally equivalent. We argue that, remarkably, experiments in which the order of two operations is conditional on a random control bit are genuine examples of circuits with cycles that have a physical realization without post-selection. The argument makes use of the general relativistic idea of background independence extended to random events. This offers a conceptual framework within which we can understand experiments where the control bit is in a quantum superposition---the so-called `quantum switch' technique \cite{chiribella3, Procopio}---as true realizations of indefinite causal structure. 

Our final step is to drop the only remnant of predefined time---the prior directionality assumed for each wire in a circuit, or, equivalently, the distinction between inputs from the past and inputs from the future. This yields a picture in which regions are connected to each other with no directionality assumed for their connections (Fig.~\ref{network}).  Each region is defined by a set of boundary systems, with the connections between regions described by entangled states that now encode the symmetry of reflection with respect to the boundary. The events taking place in a given region are represented by positive semidefinite operators on the Hilbert space of the boundary systems and can be interpreted as describing the outcomes of a measurement that the region performs on the states resulting from events in its complement. A simple rule gives the joint probabilities for the events in a network of regions (Eq.~\eqref{generalrule}).


By construction, the developed formulation of quantum theory is in agreement with observation, but time is not a fundamental concept in it. This offers the opportunity to understand time and causal structure as dynamical variables and potentially describe novel phenomena. Indeed, we discuss how the space-time metric in the familiar regimes of quantum dynamics may be possible to understand as arising from properties of the operators on the boundaries of compact space-time regions in the limit of quantum field theory, where our framework suggests a modified version of Oeckl's general boundary approach \cite{Oeckl, Oeckl3, Oeckl2}. The proposed formulation also admits more general forms of dynamics than those allowed in the standard formulation, thereby offering a framework within which to explore new physical models.  


\section*{Results}

\subsection*{The time-symmetric formulation in the circuit framework}

Consider an experiment performed in some region of space-time. It consists of a set of classical events, such as the settings of the devices used and the outcomes they produce. The events in such an experiment may be correlated with events in other experiments taking place elsewhere. Intuitively, such correlations are mediated through some information carriers, or \textit{systems}. For example, there may be correlations due to the fact that the different experiments involve measurements on systems that are correlated as a result of an event in the past, or due to the fact that some experiments take place in the past of others and there is a transfer of information from the former to the latter via certain systems. The very notion of system can be thought of as a formalization of the idea of a means through which the correlations between separate experiments are established. This notion of information exchange can be made precise in the circuit framework for operational probabilistic theories (OPTs) \cite{HardyCircuit,CDP2} (see Fig.~\ref{circuit}), of which quantum theory in its usual form can be seen as a special case. 

In Appendix A, we review the circuit framework and the standard formulation of quantum theory in it. Here, we summarize the basics of the time-symmetric formulation \cite{OC2}. Throughout this paper, we will assume finite-dimensional Hilbert spaces, and only at the end we will briefly discuss potential infinite-dimensional extensions in the context of quantum field theory.

\begin{figure}
\vspace{0.5 cm}
\begin{center}
\includegraphics[width=3cm]{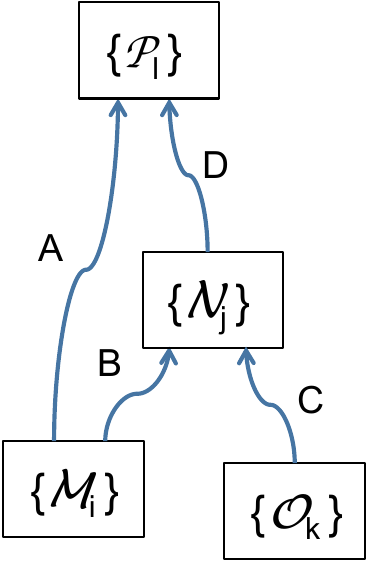}
\end{center}
\vspace{-0.5 cm}
\caption{\textbf{Standard circuit.} A standard circuit is an acyclic composition of operations with no open wires \cite{HardyCircuit, CDP2, Coecke}. An operation is a primitive type of experiment with an input and an output system (any of which could be a composite system or the trivial system), which can be thought of as performed inside an isolated box that by definition can exchange information with other operations only through the input and output systems---an idea dubbed the `closed box assumption'\cite{OC2}. Each operation has a set of possible outcomes corresponding to distinct events. In the standard approach, an operation is assumed realized without post-selection while the time-symmetric approach permits both pre- and post-selection. An OPT in the circuit framework prescribes joint probabilities for the outcomes of any given circuit (see Appendix A).} \label{circuit}
\end{figure}


An operation in the standard formulation of quantum theory is implicitly assumed realized without post-selection. In the time-symmetric formulation, both pre- and post-selection are allowed. Thus, a quantum operation from an input system $A$ to an output system $B$ is described by a collection of completely positive (CP) maps $\{\mathcal{M}^{A\rightarrow B}_j \}_{j\in O}$ corresponding to the possible outcomes ${j\in O}$, whose sum $\overline{\mathcal{M}}^{A\rightarrow B}= \sum_{i\in O}\mathcal{M}^{A\rightarrow B}_i$ is not necessarily a CP and trace-preserving (CPTP) map as in the standard formulation, but only satisfies the normalization $\textrm{Tr}(\overline{\mathcal{M}}^{A\rightarrow B}(\frac{\id^A}{d_A}))=1$. The operation resulting from the sequential composition of two operations, $\{\mathcal{L}^{A\rightarrow C}_{ij}\}_{i\in O,j\in Q} = \{\mathcal{N}^{B\rightarrow C}_j\}_{j\in Q}\circ \{\mathcal{M}^{A\rightarrow B}_i\}_{i\in O}$, has CP maps
 \begin{gather}
 \mathcal{L}^{A\rightarrow C}_{ji} =  \frac{ \mathcal{N}^{B\rightarrow C}_{j} \circ\mathcal{M}^{A\rightarrow B}_{i}} { \tr (\overline{\mathcal{N}}^{B\rightarrow C}\circ \overline{\mathcal{M}}^{A\rightarrow B}(\frac{\id^A}{d_A}))}, \hspace{0.2cm} i\in O, j \in Q,\label{seqcomp}
 \end{gather}
unless $\overline{\mathcal{N}}^{B\rightarrow C}\circ \overline{\mathcal{M}}^{A\rightarrow B} = 0^{A\rightarrow C} $, where $0^{A\rightarrow C}$ is the null CP map from $A$ to $C$. In the latter case, the composition is defined as the null operation $\{ 0^{A\rightarrow C} \}$, which is interpreted as the fact that the composition can never occur. As in the standard formulation, CP maps from the trivial system $I$ to itself are interpreted as probabilities, which implies the probability rule for all possible circuits. It is important to note, however, that even though we describe operations as collections of CP maps, which we find intuitive in view of the standard formulation, the transformation associated with a given outcome $i$ of an operation $\{\mathcal{M}^{A\rightarrow B}_j \}_{j\in O}$, defined operationally as an equivalence class of events, is not described by the CP map $\mathcal{M}^{A\rightarrow B}_i$ but by the pair of CP maps $(\mathcal{M}^{A\rightarrow B}_i; \overline{\mathcal{M}}^{A\rightarrow B})$. In particular, in the canonical representation of preparations and measurements via operators in the same space (see Appendix A), {states} are represented by $(\rho^A;\overline{\rho}^A)$, $\rho^A\leq \overline{\rho}^A$, $\tr(\overline{\rho}^A)=1$, $\rho^A, \overline{\rho}^A \in \mathcal{L}(\mathcal{H}^A)$, and {effects} by $(E^A;\overline{E}^A)$, $E^A\leq \overline{E}^A$, $\tr(\overline{E}^A)=d_A$, $E^A, \overline{E}^A \in \mathcal{L}(\mathcal{H}^A)$, with the main probability rule reading
\begin{gather}
p\left((\rho^A;\overline{\rho}^A), (E^A;\overline{E}^A)\right ) =
\frac{\tr ( \rho^A E^A)} { \tr (\overline{\rho}^A  \overline{E}^A)},\hspace{0.2cm} \textrm{for} \hspace{0.1cm}  \tr (\overline{\rho}^A  \overline{E}^A)\neq 0,\notag\\
\hspace{1.9cm} =0,\hspace{0.2cm} \textrm{for} \hspace{0.1cm}  \tr (\overline{\rho}^A  \overline{E}^A)= 0,
\label{state-effect}
\end{gather}
which reduces to Born's rule for $\rho^A=\overline{\rho}^A$, $\overline{E}^A=\id^A$. In this generalized formulation, the spaces of states and effects, understood as real functions on each other via Eq.~\eqref{state-effect}, are not closed under convex combinations, since the convex combinations of these functions in general cannot be realized in agreement with the closed-box assumption \cite{OC2}. This is the case even for deterministic states, which have the form $(\overline{\rho}; \overline{\rho})$ and hence can be described by a single normalized density matrix $\overline{\rho}$ as in the standard formulation.

Formula \eqref{state-effect} in the case of  $\tr (\overline{\rho}^A  \overline{E}^A)\neq 0$ was first derived and proposed as a fundamental rule for quantum theory by Pegg, Barnett, and Jeffers \cite{PBJ}, who regarded it as yielding a more symmetric but equivalent formulation of quantum theory. However, it is important to recognize that this formulation of quantum theory is strictly more general than the standard theory. Indeed, according to standard quantum theory, the operations implementable without post-selection form a strictly smaller class than the full class of all possible operations---they satisfy the property of \textit{causality} \cite{Pegg, CDP2,CDP}, which says that for any pair of preparation and measurement connected to each other, the probabilities of the preparation outcomes do not depend on the measurement. This restriction on pre-selected operations, which does not hold under time reversal, is not implied by the time-symmetric theory \cite{OC2}. Instead, in the time-symmetric theory it is understood as a result of asymmetric boundary conditions on the dynamics of the universe, which are also linked to the fact that we can have memory of the past but not of the future \cite{OC2}. 

The time-symmetric formulation yields an empirically consistent operational definition of time-reversal symmetry in quantum mechanics, which had been lacking in the standard formulation \cite{OC2}. While resolving this problem, it also implies the in-principle possibility for more general symmetry representations than those previously considered \cite{Wigner}. Time reversal, in particular, could have the following generalized form, which will appear naturally in the formalism that we obtain later. Let $\textrm{St}$ and $\textrm{Eff}$ denote respectively the spaces of states and effects on a system with Hilbert space $\mathcal{H}$. Under time reversal $\mathcal{T}$, every operation $\{\mathcal{M}^{A\rightarrow B}_j \}_{j\in O}$ becomes some operation  $\{\tilde{\mathcal{M}}^{B\rightarrow A}_j \}_{j\in O}$, and in particular preparations become measurements and vice versa, so states and effects get interchanged. Since $\textrm{St}$ and $\textrm{Eff}$ are separate spaces, $\mathcal{T}$ can be thought of as consisting of two maps: $\mathcal{S}_{s\rightarrow e}: \textrm{St}\rightarrow \textrm{Eff}$ and $\mathcal{S}_{e\rightarrow s}: \textrm{Eff}\rightarrow \textrm{St}$. The representation of these maps arising from the canonical representation of states and effects by pairs of operators on $\mathcal{H}$ will be denoted by $\hat{S}^A_{s\rightarrow e}$, $\hat{S}^A_{e\rightarrow s}$. The most general form of these maps is then \cite{OC2}
\begin{gather}
\hat{S}_{s\rightarrow e}( \rho; \overline{\rho}) = (F; \overline{F})= (d\frac{S\rho^T S^{\dagger}}{\tr(S\overline{\rho}^T S^{\dagger})}; d \frac{S\overline{\rho}^T S^{\dagger}}{\tr(S\overline{\rho}^T S^{\dagger})}     ), \label{Th3} \\
\hat{S}_{e\rightarrow s}( E; \overline{E}) = (\sigma; \overline{\sigma}) = (\frac{{S^{-1}}^{\dagger}E^T S^{-1}}{\tr({S^{-1}}^{\dagger}\overline{E}^T S^{-1})}; \frac{{S^{-1}}^{\dagger}\overline{E}^T S^{-1}}{\tr({S^{-1}}^{\dagger}\overline{E}^T S^{-1})}  ), \label{Th4}
\end{gather}
where $d$ is the dimension of $\mathcal{H}$, the superscript $T$ denotes transposition in some basis, and $S$ is an invertible operator on $\mathcal{H}$ which must satisfy $S = \pm {S}^T $ since time-reversal is an involution. The standard form of time reversal has this form with $S$ being unitary, which amounts to an antiunitary transformation on $\mathcal{H}$.

Finally, we remark that since in this approach an operation merely describes knowledge of the events between an input and an output system conditional on local information, operations can be updated upon learning or discarding of such information, in agreement with the corresponding Bayesian update of the joint probabilities in the circuit. The most general update rule has the form 
\begin{gather}
\{ \mathcal{M}^{A\rightarrow B}_i\}_{i\in O} \rightarrow \{ \mathcal{M}'^{A\rightarrow B}_j\}_{j\in Q},
\end{gather}
where
\begin{gather}\label{operationupdate}
\mathcal{M}'^{A\rightarrow B}_j=  \frac{\sum_{i\in O} T(j,i)\mathcal{M}_i^{A\rightarrow B}} {\sum_{j\in Q} \sum_{i\in O} T(j,i)\tr(\mathcal{M}_i^{A\rightarrow B}(\frac{\id^A}{d_A}))}\hspace{0.2cm},
\end{gather}
 \begin{gather}
 T(j,i)\geq 0, \hspace{0.2cm}\forall i\in O, \forall j \in Q, \hspace{0.2cm}\sum_{j\in Q} T(j,i)\leq 1, \forall i\in O.
\end{gather}
For example, the case in which the outcome of the operation is learned to be $i^*$ corresponds to $Q=\{i^*\}$ and $T(i^*,i) = \delta_{i^*i}$.

\subsection*{A time-neutral formulation} \label{TimeNeutrall}

Although in the time-symmetric formulation each operation can be viewed as a valid operation in either direction of time, calculating probabilities requires one to foliate a circuit and apply transformations in a particular order. We now introduce a time-neutral formulation in which we do not need to respect such an order. 

To this end, we will represent each transformation by a pair of positive semidefinite operators via a mapping inspired by the Choi-Jamio{\l}kowski (CJ) isomorphism \cite{jam,choi}, which, however, only in a special case reduces to applying a version of the Choi isomorphism to each CP map in the pair of CP maps that describes a transformation. Consider a transformation $({\cal M}^{A_1\rightarrow B_1}; \overline{{\cal M}}^{A_1\rightarrow B_1})$ (the purpose of introducing the subscript $1$ to the labels of the systems will become clear below). We will define a representation of this transformation
\begin{gather}
({\cal M}^{A_1\rightarrow B_1}; \overline{{\cal M}}^{A_1\rightarrow B_1})\leftrightarrow (M^{A_1B_2}; \overline{M}^{A_1B_2})
\end{gather}
in terms of positive semidefinite operators
\begin{gather}
M^{A_1B_2}\in{\cal L}({\cal H}^{A_1}\otimes{\cal H}^{B_2}), \hspace{0.1cm}\overline{M}^{A_1B_2} \in{\cal L}({\cal H}^{A_1}\otimes{\cal H}^{B_2}),
\end{gather}
where ${\cal H}^{B_2}$ is a copy of ${\cal H}^{B_1}$.

Take the transformation that describes time reversal for the system $B_1$ (Eqs.~\eqref{Th3}, \eqref{Th4}). Introduce a system  ${\cal H}^{A_1'}$ that is a copy of ${\cal H}^{A_1}$, and define the maximally entangled state $|\Phi^+\ket^{A_1A_1'}=\sum_{i=1}^{d_{A_1}}{|i\ket^{A_1}|i\ket^{A_1'}}/{\sqrt{d_{A_1}}} \in {\cal H}^{A_1}\otimes{\cal H}^{A_1'}$, where the set of states $\left\{|i\ket^{A_1}\right\}_{i=1}^{d_{A_1}}$ is an arbitrary orthonormal basis of  ${\cal H}^{A_1}$, and $\left\{|i\ket^{A_1'}\right\}_{i=1}^{d_{A_1}}$ is its copy in ${\cal H}^{A_1'}$. The operators $(M^{A_1B_2}; \overline{M}^{A_1B_2})$ are given by:
\begin{gather}
M^{A_1B_2}:=d_{A_1}d_{B_2}\frac{S^{B_2}\left[{\cal M}^{A_1'\rightarrow B_2}\left(|\Phi^+\ket\bra \Phi^+|^{A_1A_1'}\right)\right]^{\mathrm T}{S^{B_2}}^{\dagger}}{ \tr \left\{ S^{B_2}\left[\overline{{\cal M}}^{A_1'\rightarrow B_2}\left(|\Phi^+\ket\bra \Phi^+|^{A_1A_1'}\right)\right]^{\mathrm T}{S^{B_2}}^{\dagger}            \right\}   }, \label{CJ1}\\
\overline{M}^{A_1B_2}:=d_{A_1}d_{B_2}\frac{S^{B_2}\left[\overline{{\cal M}}^{A_1'\rightarrow B_2}\left(|\Phi^+\ket\bra \Phi^+|^{A_1A_1'}\right)\right]^{\mathrm T}{S^{B_2}}^{\dagger}}{ \tr \left\{ S^{B_2}\left[\overline{{\cal M}}^{A_1'\rightarrow B_2}\left(|\Phi^+\ket\bra \Phi^+|^{A_1A_1'}\right)\right]^{\mathrm T}{S^{B_2}}^{\dagger}            \right\}  },\label{CJ11}
\end{gather}
where $S^{B_2}$ is a copy of the operator $S^{B_1}$ that appears in the definition of time reversal for system $B_1$, and $\textrm{T}$ denotes transposition in the basis $\left\{|i\ket^{A_1}\right\}_{i=1}^{d_{A_1}}$ of $A_1$ and the basis of $B_2$ which is a copy of the transposition basis for the time reversal of $B_1$. (Every map or operator specified only on a subset of all systems is implicitly assumed extended to all systems via a tensor product with the identity on the rest of the systems, e.g., ${\cal M}^{A_1'\rightarrow B_2}\equiv \mathcal{I}^{A_1\rightarrow A_1}\otimes {\cal M}^{A_1'\rightarrow B_2}$.)

Reversely, in terms of $(M^{A_1B_2}; \overline{M}^{A_1B_2})$, the result of applying the transformation $({\cal M}^{A_1\rightarrow B_1}; \overline{{\cal M}}^{A_1\rightarrow B_1})$ on a state $(\rho^{A_1}; \overline{\rho}^{A_1})$ is given by
\begin{gather}
(\mathcal{M}^{A_1\rightarrow B_1};\overline{\mathcal{M}}^{A_1\rightarrow B_1})\circ (\rho^{A_1}; \overline{\rho}^{A_1})=(\kappa^{B_1}; \overline{\kappa}^{B_1}),
\end{gather}
\begin{gather}
\kappa^{B_1} = \frac{\tr_{A_1B_2}[ M^{A_1B_2}(\rho^{A_1}\otimes |\Phi\ket\bra \Phi|^{B_1B_2})]}{\tr[ \overline{M}^{A_1B_2}(\overline{\rho}^{A_1}\otimes |\Phi\ket\bra \Phi|^{B_1B_2})] },\label{CJ2}\\
\overline{\kappa}^{B_1} = \frac{\tr_{A_1B_2}[ \overline{M}^{A_1B_2}(\overline{\rho}^{A_1}\otimes |\Phi\ket\bra \Phi|^{B_1B_2})]}{\tr[ \overline{M}^{A_1B_2}(\overline{\rho}^{A_1}\otimes |\Phi\ket\bra \Phi|^{B_1B_2})] },\label{CJ22}
\end{gather}
where $\tr_{A_1B_2}$ denotes partial trace over $A_1B_2$, and
\begin{gather}
|\Phi\ket\bra \Phi|^{B_1B_2} = \frac{{{S^{B_2}}^{-1}}^{\dagger}|\Phi^+\ket\bra \Phi^+|^{B_1B_2} {{S^{B_2}}^{-1}}}{\tr( {{S^{B_2}}^{-1}}^{\dagger}|\Phi^+\ket\bra \Phi^+|^{B_1B_2} {{S^{B_2}}^{-1}}   ) },
\end{gather}
with $|\Phi^+\ket^{B_1B_2} =\sum_{i=1}^{d_{B_1}}{|i\ket^{B_1}|i\ket^{B_2}}/{\sqrt{d_{B_1}}} \in {\cal H}^{B_1}\otimes{\cal H}^{B_2}$, where
$\left\{|i\ket^{B_1}\right\}_{i=1}^{d_{B_1}}$ is the transposition basis for the time reversal of $B_1$, and $\left\{|i\ket^{B_2}\right\}_{i=1}^{d_{B_1}}$ is its copy in $B_2$. (Here, again, tensor product with the identity is implicit, e.g., $M^{A_1B_2}\equiv\id^{B_1}\otimes M^{A_1B_2}$.) Since $S^{B_1}$ is defined up to an overall factor, without loss of generality we will assume that it is normalized as
\begin{gather}
\tr({{S^{B_1}}^{-1}}^{\dagger}{{S^{B_1}}^{-1}})= d^{B_1},
\end{gather}
and we will simply write
\begin{gather}
|\Phi\ket^{B_1B_2} ={{S^{B_2}}^{-1}}^{\dagger}|\Phi^+\ket^{B_1B_2}.
\end{gather}
Furthermore, since time reversal is an involution, we have $S^{B_1} = {S^{B_1}}^T $ or $S^{B_1} =- {S^{B_1}}^T $ (corresponding to the form of time reversal for bosons and fermions, respectively, in the case when $S^{B_1}$ is unitary \cite{Wick}), and this implies that the vector $|\Phi\ket^{B_1B_2}$ is either symmetric or anti-symmetric with respect to interchanging systems $B_1$ and $B_2$,
\begin{gather}
|\Phi\ket^{B_1B_2} = \pm {{S^{B_1}}^{-1}}^{\dagger}|\Phi^+\ket^{B_1B_2} .
\end{gather}
This (anti-)symmetry of the state makes sense only if we have a correspondence between the basis of $B_1$ and the basis of $B_2$, which was assumed here. As we will see, the physical meaning of this correspondence is given precisely by the transformation of time reversal.

At the level of operators, the overall sign disappears, and we simply have 
\begin{eqnarray}
|\Phi\ket\bra \Phi|^{B_1B_2} &&= {{{S^{B_1}}^{-1}}^{\dagger}|\Phi^+\ket\bra \Phi^+|^{B_1B_2} {{S^{B_1}}^{-1}}}\notag\\
&&= {{{S^{B_2}}^{-1}}^{\dagger}|\Phi^+\ket\bra \Phi^+|^{B_1B_2} {{S^{B_2}}^{-1}}}.
\end{eqnarray}

 In the above representation, a general operation $\{\mathcal{M}^{A_1\rightarrow B_1}_i\}_{i\in O}$ can be described by a collection of positive semidefinite operators $\{M^{A_1B_2}_i\}_{i\in O}$ with the normalization $\tr( \sum_{i\in O} M^{A_1B_2}_i)\equiv \tr( \overline{M}^{A_1B_2}) = d_{A_1}d_{B_2}$. In the case when $S=\id$, the representation agrees with a Choi isomorphism \cite{choi} (modulo an overall transposition \cite{OCB} and a different normalization) applied to each CP map in the operation. 


In order to represent the formula for the probabilities of a circuit in a way that does not require writing down the operations in any particular order, we will describe the CP maps in the separate boxes as operators defined on \textit{separate} systems \cite{OCB}, even if the boxes may be connected to each other by wires. Figuratively, one can think that each junction where a wire gets attached to a box in the graphical representation of operations corresponds to a different system. Each wire will therefore be associated with two systems---one for each end of the wire---instead of just a single system. It is important to note that the objects in a circuit are not assumed to have specific temporal durations; they are simply logical transformations. But if we nevertheless think that there is a time duration associated with them in a particular implementation, a wire is supposed to have a zero duration---it merely represents the connection between operations and would be associated with the instant at which one operation ends and another one begins. Therefore, the fact that we propose to associate two systems with each wire should not be confused with the idea that these are systems associated with different instants of time---the two systems corresponding to the ends of the same wire are associated with the same instant.

\begin{figure}
\vspace{0.5 cm}
\begin{center}
\includegraphics[width=8cm]{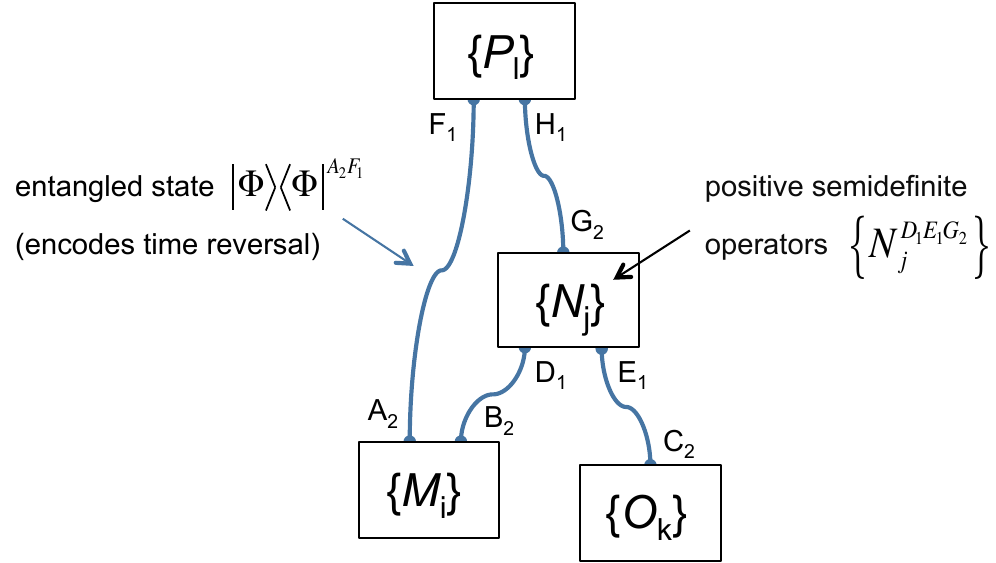}
\end{center}
\vspace{-0.5 cm}
\caption{\textbf{Time-neutral circuit formulation.} With each wire in a circuit we associate two Hilbert spaces---one for each end of the wire. The Hilbert space associated with the `future' end of a wire is referred to as of type $1$, while the one associated with the `past' end of a wire as of type $2$. Each wire is described by some entangled state $|\Phi\rangle\langle\Phi|$ on the tensor product of the two Hilbert spaces. The operations in the boxes are described by collections of positive semidefinite operators on the Hilbert spaces of the wire ends attached to it. The probabilities for the events in the circuit are given by Eq.~\eqref{Wcircuit}. } \label{twosyst}
\end{figure}

We will label the different systems by capital letters, $A$, $B$, $C$, etc., with the subscript $1$ added to those system that are attached to the past side of a box (each of these corresponds to the `future' end of some wire), and the subscript $2$ added to those systems that are attached to the future side of a box (these correspond to the `past' end of some wire); see Fig.~\ref{twosyst}. We will refer to these systems as systems of type $1$ and type $2$, respectively. 
The operations inside each box will be represented by collections of positive semidefinite operators according to the representation above, defined on the Hilbert spaces of the systems attached to the box, whose subscripts $1$ and $2$ have been chosen to match those in the definition. We will use different letters to denote the operators of events in different boxes. The different operations will thus be written as collections of operators $\{M^{A_2B_2}_i\}_{i\in O}$, $\{N^{D_1E_1G_2}_j\}_{j\in Q}$, etc.

With each wire in the circuit, such as the one between systems $A_2$ and $F_1$ in Fig.~\ref{twosyst}, we will associate the entangled state $W^{A_2F_1}\equiv|\Phi\rangle\langle\Phi|^{A_2F_1}$, defined as explained above, and with the collection of all wires in the circuit, we will associate the tensor product of the corresponding states, $W^{A_2F_1B_2D_1\cdots}=|\Phi\rangle\langle\Phi|^{A_2F_1} \otimes |\Phi\rangle\langle\Phi|^{B_2D_1}\otimes \cdots$. The joint probabilities for the outcomes of a set of operations connected in a circuit, such as the one depicted in Fig.~\ref{twosyst}, are then given by
\begin{gather}
p(i,j,\cdots |\{M^{A_2B_2}_i\}_{i\in O}, \{N^{D_1E_1G_2}_j\}_{j\in Q},\cdots, \textrm{circuit})=\notag\\
\frac{\tr\left[W^{A_2F_1B_2D_1\cdots} \left(M^{A_2B_2}_i\otimes N^{D_1E_1G_2}_j \otimes\cdots\right)\right]  }{ \tr \left[W^{A_2F_1B_2D_1\cdots} \left(\overline{M}^{A_2B_2}\otimes \overline{N}^{D_1E_1G_2} \otimes\cdots\right)\right]}, \label{Wcircuit}
\end{gather}
for ${ \tr \left[W^{A_2F_1B_2D_1\cdots} \left(\overline{M}^{A_2B_2}\otimes \overline{N}^{D_1E_1G_2} \otimes\cdots\right)\right]}\neq 0$, or 
\begin{gather}
p(i,j,\cdots |\{M^{A_2B_2}_i\}_{i\in O}, \{N^{D_1E_1G_2}_j\}_{j\in Q},\cdots, \textrm{circuit})=0, \label{Wcircuit2}
\end{gather}
for ${ \tr \left[W^{A_2F_1B_2D_1\cdots} \left(\overline{M}^{A_2B_2}\otimes \overline{N}^{D_1E_1G_2} \otimes\cdots\right)\right]}= 0$.

The validity of this formula can be verified easily from the isomorphism equations \eqref{CJ1}, \eqref{CJ11} and \eqref{CJ2}, \eqref{CJ22}. Indeed, from \eqref{CJ1}, \eqref{CJ11} we see that the operators describing preparation events coincide with the time-reversed images of the corresponding states. By contracting (taking the partial trace of) these operators with the state $|\Phi\rangle\langle\Phi|$ of the wire attached to the box, we obtain the actual states on the other end of the wire, which is an input of a subsequent box.  From \eqref{CJ2}, \eqref{CJ22} we see that by contracting the operators describing the transformations in any box with the state on its input systems and with the state of the output wires on those ends attached to the box, leaves on the other end of the output wires the result of the transformation in that box applied on the input state. These transformations continue until the evolved state is finally contracted with the effect of a final measurement, which in this representation is identical to the representation $(E,\overline{E})$ used earlier. 

Notice that if we imagine disconnecting a wire in the circuit at a junction corresponding to a system of type $1$, the pair of operators that we obtain on this end of the wire by contracting the events in the boxes and the wires in the past, is exactly the state that would result from the sequence of past transformations up to that point. In this sense, the notion of system in the usual formulation of quantum theory corresponds to the systems of type $1$. Similarly, the states that we obtain under time-reversal \eqref{Th4} live on the systems of type $2$.

Now we can understand the meaning of the isomorphism between the Hilbert spaces on the two ends of a wire that was assumed by declaring that they are copies of each other. Consider a preparation and a measurement connected by a wire, where the preparation box is attached to the system of type $2$ and the measurement box to the system of type $1$. Imagine that we could take the measurement box and physically `flip' its time orientation so that we could plug it in the place of the preparation box, that is, attach it to the type-$2$ end of the wire (in practice, this would mean to create a preparation box that looks just like the measurement box operating in reverse order). This flipped box now gives rise to a preparation that can be operationally characterized by measurement boxes connected to the type-$1$ end of the wire, which have not been flipped. By definition, this preparation is the time-reversed image of the original measurement. The states that we obtain on the system of type $2$ when we connect the original (not flipped) measurement to the type-$1$ end of the wire, are copies of the states that we would obtain on the type-$1$ end of the wire if we attach the flipped measurement to the type-$2$ end. The equivalence between the different states on both ends of the wire is therefore exactly the one defined by the physical transformation of time reversal.

Although we defined the descriptions of operations and wires assuming the transformation of time reversal, this should be regarded as reverse engineering. What we propose is that quantum theory is described by the general formalism above, where the content of each box is represented by a set of positive semidefinite operators and each wire is described by some entangled pure state $|\Phi\rangle\langle\Phi|$. \textit{A priori}, the states of the wires can be any, and which specific states they are is a feature of the mechanics that we find out to govern the physics around us. Thus, time reversal should be understood as determined by the states of the wires, rather than the other way around. (The entangled state of a wire can be assumed to have a maximal Schmidt rank, since if it does not, we can redefine the dimension of the systems on both ends of the wire.) Even though time reversal as understood at present is described by a unitary $S$ (corresponding to maximally entangled states of the wires), we will leave open the possibility for arbitrary $S$, since nothing in the theory gives reasons to discard it.

\subsection*{Circuits with cycles and the process operator}

\begin{figure}
\vspace{0.5 cm}
\begin{center}
\includegraphics[width=3.5cm]{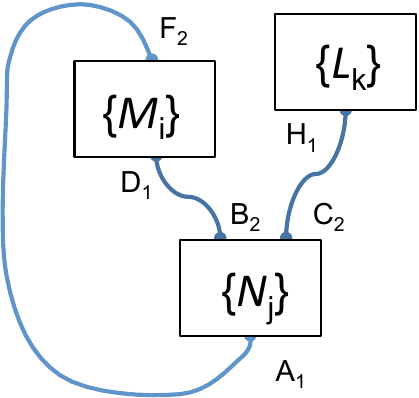}
\end{center}
\vspace{-0.5 cm}
\caption{\textbf{Cyclic circuit.} Formula \eqref{Wcircuit} can be applied to circuits with cycles too.} \label{loop}
\end{figure}

The form of Eq.~\eqref{Wcircuit} treats the information about the wiring between boxes separately from the information about the content of the boxes. This allows us to extend the framework to circuits that involve cycles, such as the one in Fig.~\ref{loop}. We simply define the same formula \eqref{Wcircuit} to provide the probabilities in such cases too, with $W$ encoding the corresponding wiring. This rule is in agreement with a model of quantum theory in the presence of closed time-like curves (CTCs) \cite{Hartle, HorMald, GotPre, Bennett2, Svetlichny, Lloyd, Bennett1, CTC, daSilva, GCH} which has become known as post-selected CTCs, since it can be simulated by post-selection.

\begin{figure}
\vspace{0.5 cm}
\begin{center}
\includegraphics[width=5cm]{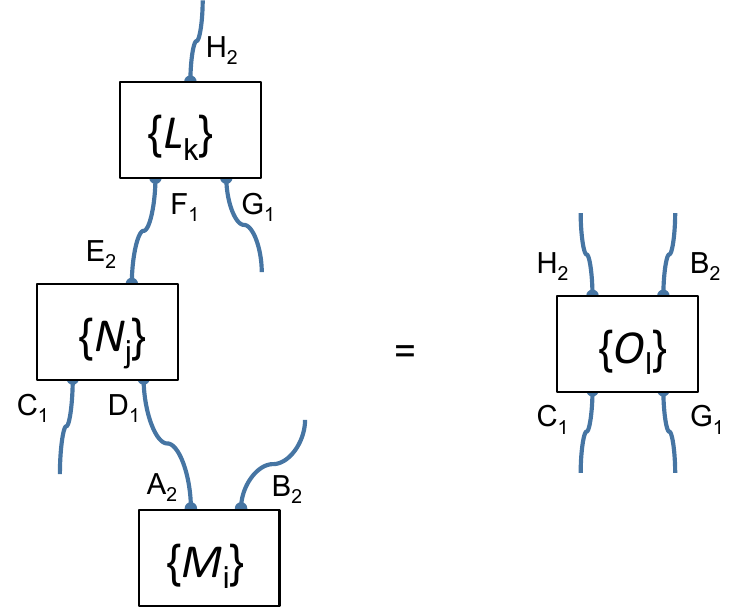}
\end{center}
\vspace{-0.5 cm}
\caption{\textbf{Fragment of a circuit.} Within the generalized circuit framework that permits cycles, any fragment of a circuit is a valid operation.} \label{segment}
\end{figure}

In the more general paradigm permitting cycles, any fragment of an acyclic circuit, such as the one in Fig.~\ref{segment}, can be regarded as a valid operation if we disregard the underlying causal structure with respect to which the different inputs and outputs of the fragment are ordered, and regard the circuit as a cyclic one. Indeed, since all systems of type $1$ are always connected to systems of type $2$, the fragment and its complement can be seen as two operations connected to each other in a loop. A formalism that describes fragments of standard circuits, called \textit{quantum combs}, has been developed by Chiribella, D'Ariano, and Perinotti \cite{networks}. An alternative formalism, the \textit{duotensor} framework, has been developed by Hardy  \cite{Hardy2}. In our formalism, the hierarchy of quantum combs collapses and all fragments are equivalent to operations.

Consider the composition between two operations $\{ M^{A_1B_1C_2D_2}_i\}_{i\in O}$ and $\{ N^{E_1F_1G_2H_2}_j\}_{j\in Q}$, where system $A_1$ is connected to system $G_2$ and system $E_1$ is connected to system $C_2$ (the systems in each pair obviously must have the same dimension). The resultant operation is  $\{L^{B_1F_1D_2H_2}_{ij}\}_{i\in O, j\in Q}$, where
\begin{widetext}
\begin{gather}
L^{B_1F_1D_2H_2}_{ij}
 = {d_{B_1}d_{F_1}d_{D_2}d_{H_2}} \frac{\tr_{A_1G_2E_1C_2}\left[|\Phi\rangle\langle\Phi|^{A_1G_2}\otimes  |\Phi\rangle\langle\Phi|^{E_1C_2}  \left( M^{A_1B_1C_2D_2}_i\otimes N^{E_1F_1G_2H_2}_j\ \right) \right] } {\tr \left[ |\Phi\rangle\langle\Phi|^{A_1G_2}\otimes  |\Phi\rangle\langle\Phi|^{E_1C_2}  \left( \overline{M}^{A_1B_1C_2D_2}\otimes \overline{N}^{E_1F_1G_2H_2}\ \right) \right]  }, \hspace{0.2cm}
 \forall i\in O, j\in Q,
\end{gather}
\end{widetext}
for ${\tr \left[ |\Phi\rangle\langle\Phi|^{A_1G_2}\otimes  |\Phi\rangle\langle\Phi|^{E_1C_2}  \left( \overline{M}^{A_1B_1C_2D_2}\otimes \overline{N}^{E_1F_1G_2H_2}\ \right) \right]  }\neq 0$, or the null operation $\{0^{B_1F_1D_2H_2}\}$ otherwise. This formula expresses the most general composition rule in this framework, because, without loss of generality, all systems of type $1$ for each of the two operations can be grouped into two systems, one of which is being connected while the other one is left free, and similarly for the systems of type $2$. It also captures the notion of parallel composition, which can be thought of as the case where the trivial (1-dimensional) system of type $2$ of one operation is connected to the trivial input system of type $1$ of another. The compositions of three or more operations  also follows from this rule. Similarly to the case of acyclic circuits, the probabilities in the formula \eqref{Wcircuit} for general circuits with cycles can be seen as the operators associated with the outcomes of an operation from the trivial system to itself.

From formula \eqref{Wcircuit}, we can obtain a similar expression for the joint probabilities of the outcomes of only a proper subset $\{M^{A_1B_2}_i\}_{i\in O}, \{N^{C_1D_2}_j\}_{j\in Q},\cdots$ of all operations in a circuit, conditionally on information about the events in the rest of the circuit. Let us denote the variables describing the rest of the circuit collectively by $w$.
Then the probabilities have the form
\begin{gather}
p(i,j, \cdots|\{M^{A_1B_2}_i\}_{i\in O}, \{N^{C_1D_2}_j\}_{j\in Q},\cdots; w)\notag\\
=\frac{\tr\left[ W^{A_1B_2C_1D_2\cdots}\left(M^{A_1B_2}_i\otimes N^{C_1D_2}_j\otimes\cdots\right)\right] } {\tr\left[ W^{A_1B_2C_1D_2}\left(\overline{M}^{A_1B_2}\otimes \overline{N}^{C_1D_2}\otimes\cdots\right) \right]}, \label{pmform}
\end{gather}
where most generally
\begin{gather}
W^{A_1B_2C_1D_2\cdots}\geq 0, \hspace{0.2cm}\tr W^{A_1B_2C_1D_2\cdots}=1.\label{procond}
\end{gather}
(Again, for ${\tr\left[ W^{A_1B_2C_1D_2}\left(\overline{M}^{A_1B_2}\otimes \overline{N}^{C_1D_2}\otimes\cdots\right) \right]}= 0$, the probabilities are defined to be zero.) 

In Appendix B, we show that any operator satisfying Eq.~\eqref{procond} can be obtained in practice by embedding the separate operations in a suitable acyclic circuit with operations that may involve post-selection. We show that this can also be interpreted as embedding the separate operations in a circuit with a cycle, and that any cyclic circuit has a physical realization in this way. We also argue that any physical circumstances in which the correlations between a set of separate operations is described by expression \eqref{pmform}, even if they are not assumed to arise from a circuit, can be interpreted as a circuit with a cycle. This is illustrated in the following section. 

The operator $W$ in expression \eqref{procond} is a generalization of the \textit{process matrix} introduced in Ref.~\cite{OCB}. We will refer to it in the same way, or more precisely, as the \textit{process operator}, since the matrix is its description in a given basis. The original concept was proposed as a means of describing the correlations between local standard quantum operations without the assumption that the operations are part of a standard quantum circuit. It was derived assuming that the joint probabilities for the outcomes of the operations are non-contextual and linear functions of the CP maps describing their outcomes, as well as that the local operations can be extended to act on joint input ancillas in arbitrary quantum states. The requirement of linearity and normalization on local CPTP maps in that approach gives rise to additional asymmetric constraints on $W$, which are not part of the present framework.

Remarkably, in the present framework the process operator can be understood as the operator describing a deterministic quantum state. Note that formula \eqref{pmform} is exactly analogous to the formula for the probabilities for the outcomes of a set of local measurements of the generalized kind applied on a joint deterministic state $W$. Ignoring the subscripts $1$ and $2$, the operators of an operation $\{M^{A_1B_2}_i\}_{i\in O}$ are equivalent to the operators describing a measurement on a pair of input systems $A_1$ and $B_2$. By construction, they have the right normalization, and in the case when the system $B_2$ is trivial, $\{M^{A_1}_i\}_{i\in O}$ coincide with the standard measurement operators. The process operator $W^{A_1B_2C_1C_2\cdots}$ also has the form of a deterministic state, and in the case of trivial systems of type $2$, it coincides with a standard deterministic state. Furthermore, when the systems of type $1$ are trivial, the operators of the operations coincide with the operators of the corresponding time-reversed measurements obtained via Eq.~\eqref{Th3}, while the operator $W$ coincides with the corresponding time-reversed state. Therefore, a general operation $\{M^{A_1B_2}_i\}_{i\in O}$ can be thought of as implementing a joint destructive measurement on two input systems---one from the past and one from the future. The deterministic state $W^{A_1B_2}$ on which that measurement is applied, which generally depends on events both in the past and in the future of the measurement, is our version of the two-state vector idea of Aharonov, Bergmann, and Lebowitz (ABL) \cite{ABL} (see also Watanabe \cite{Watanabe}). In the simple special case when the operation is sandwiched between the preparation of a state $|\psi\rangle\langle\psi |$ and a post-selection on a measurement outcome with operator $|\phi\rangle\langle\phi |$, our state is $|\psi\rangle\langle\psi |^{A_1}\otimes {S^{B_2}}^{-1}{|\phi \rangle\langle\phi |^T}^{B_2}{{S^{B_2}}^{-1}}^{\dagger}$. This formally resembles the original two-state vector, but there are principal differences between the two-state vector and the state above. One difference is that the two states in the ABL formalism are associated with the same time instant, whereas here they are associated with two different times. At a single instant, we can also have two systems, but these are the two ends of a wire and they are always in the entangled state $|\Phi\rangle\langle\Phi|$. More importantly, the backward `evolving' state in the ABL two-state vector lives in the dual of the forward-oriented state space, i.e., it is actually an effect. In contrast, here the state ${S^{B_2}}^{-1}{|\phi \rangle\langle\phi |^T}^{B_2}{{S^{B_2}}^{-1}}^{\dagger}$ is the image of the effect under the physical transformation of time reversal, and it thus literally represents a state with reverse time orientation. The effects in our picture are the operators $(M^{A_1B_2}_i; \overline{M}^{A_1B_2})$ describing `transformations'. 
The operator $W^{A_1B_2C_1C_2\cdots}$ in the general case is our analogue of the most general concept developed in the two-state vector approach---the multi-time mixed state \cite{multitime}---again differing in its meaning and axiomatics as described. We note that an isomorphism between the two-state picture and the present picture, for $S=\id$ and without a physical interpretation of the second system or the choice of transposition basis, has been noticed in Ref.~\cite{Silva}.

It is important to emphasize that the state $W^{A_1B_2C_1C_2\cdots}$ is not supposed to be interpreted as a description of the events that exist over some portion of space-time. The description of events in space-time is given by a circuit, not a state. A state is associated with the free wire ends of a circuit fragment. Figuratively, we can imagine removing some of the boxes in a circuit, thus leaving certain wires free. A state is then associated with the wire ends on the boundary of the empty region, while the content of the region, that is, the box plugged in it, describes the measurement applied on that state. 

The Choi isomorphism has been previously used in frameworks describing transformations of transformations \cite{networks, supermaps,OCB}, but merely as a convenient representation. Indeed, the Choi operator of a CP map is dependent on an arbitrary choice of basis. In contrast, the isomorphism defined here is based on the physical transformation of time reversal, which supplies the formalism with a physical interpretation.  


\subsection*{Existence of cyclic circuits without post-selection} \label{Existence}

As we have seen, the paradigm of circuits that permit cycles allows us to treat acyclic circuits in a more general fashion by regarding any fragment of a circuit as a valid operation. So far, this is only a more general formalism applied to phenomena that can be understood by acyclic circuits as well. We now argue that there exist simple phenomena obtained with no post-selection, which cannot be described by acyclic circuits, but can be understood as examples of circuits with cycles according to our notion of operation.

We will assume that $S=\id$ for simplicity, since the argument does not depend on the exact form of time reversal. A simple example is a process operator for two separate operations,  $\{M^{A_1B_2}_i\}_{i\in O}$ and $\{N^{C_1D_2}_j\}_{j\in Q}$, which has the form $W^{A_1B_2C_1D_2} = \frac{1}{2} \rho^{A_1}\otimes |\Phi^+\rangle\Phi^+|^{B_2C_1}\otimes \frac{\id^{D_2}}{d_{D_2}}+ \frac{1}{2} \rho^{C_1}\otimes |\Phi^+\rangle\Phi^+|^{D_2A_1}\otimes \frac{\id^{B_2}}{d_{B_2}}$ (here $d_{A_1}=d_{B_2}=d_{C_1}=d_{D_2}$). This process operator is an equally weighted convex mixture of two process operators. One of them describes a situation where a state $\rho$ is fed into the input of the operation $\{M^{A_1B_2}_i\}_{i\in O}$, after which the output system of $\{M^{A_1B_2}_i\}_{i\in O}$ is sent through a perfect channel into the input of $\{N^{C_1D_2}_j\}_{j\in Q}$, and then the output of  $\{N^{C_1D_2}_j\}_{j\in Q}$ is discarded, i.e., subjected to the standard unit effect [in our language, the effect $(\id;\id)$]. The other process operator describes the analogous situation with the roles of $\{M^{A_1B_2}_i\}_{i\in O}$ and $\{N^{C_1D_2}_j\}_{j\in Q}$ interchanged. The process operator $W^{A_1B_2C_1D_2} $ does not correspond to the operations being embedded in a fixed standard circuit without post-selection, because for any such circuit, when the operations are standard quantum operations, there must be zero signaling in at least one direction between the operations, while here we can have some signaling in both. The correlations described by this process operator could be obtained in practice by implementing at random one of the two circuit scenarios corresponding to the two process operators of which the whole process operator is a mixture. We can imagine that this is done in such a way that the operation $\{N^{C_1D_2}_j\}_{j\in Q}$ is always applied at a fixed time, while $\{M^{A_1B_2}_i\}_{i\in O}$ may be applied before or after that time depending on which circuit scenario is realized. The time of $\{M^{A_1B_2}_i\}_{i\in O}$ can be determined conditionally on the value of a classical random bit, which we will refer to as the control bit. 

\begin{figure}
\vspace{0.5 cm}
\begin{center}
\includegraphics[width=8.5cm]{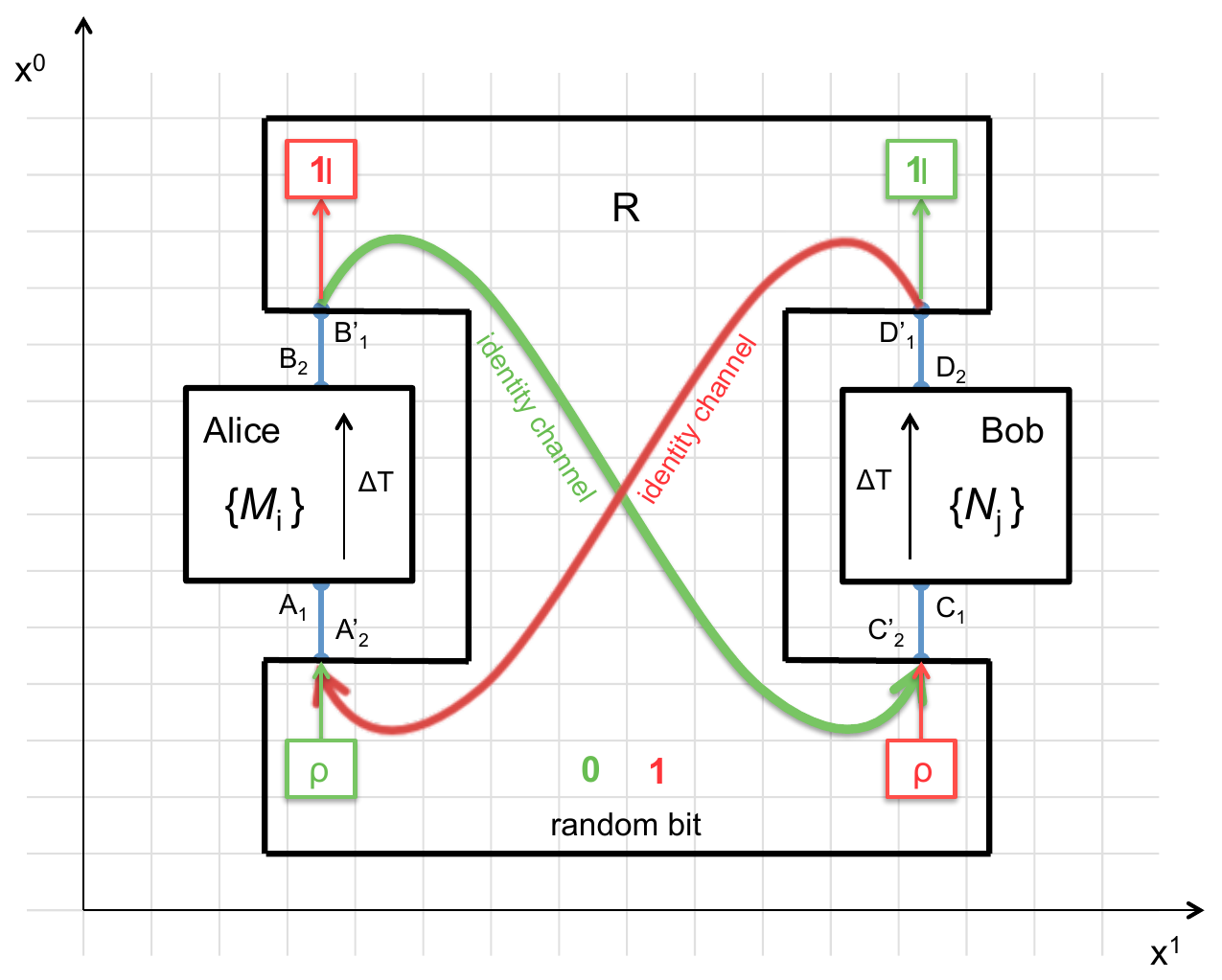}
\end{center}
\vspace{-0.5 cm}
\caption{\textbf{Cyclic circuit without post-selection.} The operation of Alice and Bob are embedded in one of two possible circuits (depicted in green and red) realized conditionally on the value of a random control bit ($0$ or $1$). Alice and Bob do not share a common time reference and each performs her/his operation upon receiving of the input system during a fixed time interval $\Delta T$. Each of them thus exchanges information with the rest of the experiment only through the respective input and output systems. When the value of the control bit is unknown, the operation $R$ taking place in the external region does not correspond to a fragment of an acyclic circuit that contains the operations of Alice and Bob. Nevertheless, each region contains a valid operation and all operations are connected in a circuit.} \label{AandB}
\end{figure}

We want to argue that in such a situation we can still think of the operation $\{M^{A_1B_2}_i\}_{i\in O}$ as applied once in agreement with the closed-box assumption, even if it may occur at two possible times. Let us imagine that the operations  $\{M^{A_1B_2}_i\}_{i\in O}$ and $\{N^{C_1D_2}_j\}_{j\in Q}$ are performed by two experimenters, Alice and Bob, respectively, each of whom resides inside a closed laboratory and applies the respective operation within a fixed time duration $\Delta T$ upon receiving of an input system, after which the transformed system is immediately sent out (the necessary transmission of information between the laboratories of Alice and Bob can be implemented by a suitable mechanism outside). The reason why we require that the operations of each party have a fixed time duration is to exclude the possibility that by modulating the duration the parties can exchange with the outside world information additional to the one carried through the input and output systems. We also need to make sure that the parties do not possess clocks synchronized with outside events in the experiment, since otherwise they could learn additional information by reading the time at which they receive their input systems. We can imagine that each of them implements her/his operation with a stop watch, which is started upon receipt of the input system and stopped upon release of the output system. The fact that Bob performs a valid operation in these circumstances should be non-controversial since his experiment is the paradigmatic example of what an operation is envisioned to correspond to in practice. Even though Alice's operation may take place at two possible times as measured by an external clock, it should be intuitively clear that from the point of view of Alice, her experiment looks no differently than the way Bob's experiment looks to Bob.

To illustrate explicitly that both Alice's and Bob's experiments are valid operations, we will analyze the setup from a more general space-time perspective. As is well known \cite{Gravitation}, in classical physics the coordinates in the space-time manifold do not have a physical meaning. When we describe physical phenomena in terms of such coordinates, we may choose any coordinate grid. The causal structure of space-time (captured by the null geodesics), when described relative to that grid, may curve and twist in any direction depending on the choice of the grid, which, however, would not represent different physics as it is only the relational degrees of freedom between physical objects that matter. We will assume that the same remains true when we analyze random events of the kind above, and will illustrate the closed-box idea in a suitable graphical representation by choosing coordinates in the space-time manifold such that the operations of Alice and Bob take place in fixed regions as described by these coordinates (Fig.~\ref{AandB}). Inside each of the two regions, we can assume a fixed causal structure where the input precedes the output. The stopwatches of Alice and Bob in those regions would display readings in a fixed range from the input to the output. However, without any additional information, the causal structure and events taking place outside of the two regions can be any. In a situation in which Alice performs her operation first, and then her output is sent to the input of Bob, the future light cone of Alice's output must curve in such a way with respect to the chosen grid that Bob's input is inside it. In such a case, the circuit diagram describing the transmission of information, curved correspondingly, would look as the green picture in Fig.~\ref{AandB}. In the case when Bob is before Alice, the circuit would look like the red picture in Fig.~\ref{AandB}.  Notice that in each of these cases, the region that connects the boxes of Alice and Bob, which contains the events relevant to this experiment that are external to these boxes, corresponds to a single operation from the outputs of the parties to their inputs. However, in each of these cases, this operation can be seen as a fragment of a standard quantum circuit that contains the operations of Alice and Bob. But when the control bit is unknown and therefore the causal structure in that intermediate region is unknown, the operation that we assign to that region does not correspond to a fragment of an acyclic circuit that contains the operations of Alice and Bob. It is nevertheless a valid operation from the point of view of the closed-box assumption [described by the operator $R^{A_2'B_1'C_2'D_1'}=d_{A_2'B_1'C_2'D_1'} (W^T)^{A_2'B_1'C_2'D_1'}$], because the information about it is obtained without looking into the boxes of Alice and Bob. We therefore see that we have a bona fide example of a nontrivial cyclic circuit, which can be realized in practice without post-selection.

This example can be readily extended to cases where the control bit is prepared in a quantum superposition. This gives rise to the so-called `quantum switch' of the operations of Alice and Bob \cite{chiribella3}, which has been shown to allow implementing certain tasks that cannot be achieved if the order of operations is called in a classically definite order \cite{chiribella3, Colnaghi, Chiribella12b, Araujo} and was recently demonstrated experimentally in a simple setup \cite{Procopio}. If in such a case the control qubit together with the output of the last party is fed into the input of a third party, Charlie, the resultant tripartite quantum process operator connecting the three parties can be shown to be causally non-separable \cite{OG} (see also Ref.~\cite{Araujo3}). From the outlined perspective, we can understand the region connecting the three parties as containing an event, which is a pure superposition of events compatible with fixed causal orders. Note that in order not to destroy the superposition, the operations of Alice and Bob should be performed in such a way that they do not leave track of the time at which the operations are performed. In practice, this may require coherent manipulation of the devices, which may be unrealistic for macroscopic devices, but is in principle compatible with quantum mechanics \cite{OG}.

A cyclic circuit does not mean transmission of information back in time relative to our usual time. While the operation in the region between Alice and Bob  in Fig.~\ref{AandB} is connected to the local operations in a loop, its input and output systems do not generally have well defined times. When they do (the case of the control bit having value exactly $0$ or exactly $1$), the two input systems $B'_1$ and $D'_1$ are associated with different times just like the output systems $A_2'$ and $C_2'$, and the information transmission respects the time ordering. For example, in the specific case depicted in green, the transmission from input to output effected by the box is localized between $B'_1$ to $C'_2$, and the time of $B'_1$ is before the time of $C'_2$. The time of $B'_1$ is after the time of $A'_2$, but there is no transmission of information from $B'_1$ to $A'_2$. 


\subsection*{Dropping the assumption of predefined time}

The fact that the situation in the previous example corresponds to a cyclic circuit without an actual transmission of information back in time illustrates the fact that the circuit structures in the more general framework we are considering are logical structures, in which the order of operations does not necessarily correspond to time. But if time is not the ordering of operations, then what is it?

As we have seen in the example of Fig.~\ref{AandB}, the causal structure outside of the regions of Alice and Bob, whenever well defined, is reflected in the form of the operation taking place in that region. This suggests that time and causal structure should be searched for in properties of the contents of the boxes in space-time, rather than the way boxes are composed. But in the framework developed so far, we have a distinction between systems of type $1$ and type $2$, which was inherited from the background time assumed in deriving the formalism. Does this mean that some primitive notion of time needs to be postulated?

Observe that within the generalized process operator formalism \eqref{pmform}, which can be interpreted as describing the outcomes of measurements on a joint quantum state, we can arbitrarily redefine the types of the different systems ($1$ or $2$, interpreted as inputs from the past and from the future, respectively), and the expression \eqref{pmform} remains valid except that we would attach different interpretations to the systems. This shows that the labels $1$ and $2$ are superfluous as far as the probabilities of events are concerned.

\begin{figure}
\vspace{0.5 cm}
\begin{center}
\includegraphics[width=6.0cm]{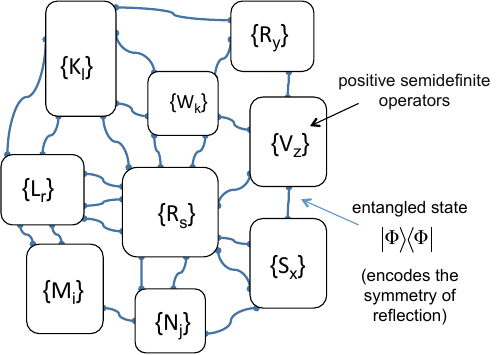}
\end{center}
\vspace{-0.5 cm}
\caption{\textbf{Network of quantum operations without assumption of predefined time.} Each operation describes knowledge about the possible events in a region conditional on local information. Regions are connected to other regions through parts of their boundaries, with the connections graphically represented by ‘wires’. Unlike the circuit framework, the wires do not have predefined directions, and in the general case (e.g., if the boundary is time-like) could transmit information in both directions. Each wire is described by a pure bipartite entangled state $|\Phi\rangle\langle\Phi|$ on the tensor product of the two Hilbert spaces associated with its two ends. The joint probabilities for a network of operations is given by Eq.~\eqref{generalrule}.} \label{network}
\end{figure}

Our proposal therefore is to abandon the \textit{a priori} distinction between systems of type $1$ and type $2$ and regard all systems as equivalent. A further rationale for this comes from conceiving the quantum field theory limit of the framework. Within a given precision, we could think that the dynamics of the quantum fields in space-time is approximated by a discrete unitary circuit on a lattice. A given region of space-time would cut out a fragment of the circuit and its boundary would be pierced by many input and output wires whose distribution depends on the shape of the region. If the region has a suitable shape, such as a lens enclosed by two space-like hypersurfaces, the fragment can be equivalent to a standard operation where all input wires are in the past of all output wires. In such a case, we are allowed to group all input wires and regard them as a single wire of a larger dimension associated with the input hypersurface, and similarly for the output wires. The division into fundamental discrete wires would disappear in the continuum limit, suggesting that a wire is to be associated with a local area on the boundary hypersurface. However, if the boundary contains time-like parts, the `wire' corresponding to a time-like area (composed of many input and output wires in the discrete approximation) would generally transmit information in both directions, since information can flow both into and out of the region through such an area. This shows that if we think of a wire as a local area of contact between two regions, it need not be associated with any particular directionality.

The OPT that we obtain by dropping the distinction between systems of type $1$ and $2$ is summarized by the following rules.\\

(1) An operation is a set of events $\{M^{AB\cdots}_i\}_{i\in O}$ in a region defined by some boundary systems $A$, $B$, $\cdots$, associated with Hilbert spaces $\mathcal{H}^A$, $\mathcal{H}^B$, $\cdots$, of dimension $d^A$, $d^B$, $\cdots$. The events are described by positive semidefinite operators $M^{AB\cdots}_i\geq 0$ on $\mathcal{H}^A\otimes \mathcal{H}^B \otimes \cdots$ with the normalization $\tr(\sum_{i\in O} M_i) \equiv \tr \overline{M} = d^Ad^B\cdots$. An exception is the null operation $\{ 0^{AB\cdots}\}$, which is a singular case. \\

(2) Two operations may be connected through some of their boundary systems whenever these systems are of the same dimension. Such a connection, pictorially represented by a wire connecting the regions, is associated with a bipartite pure entangled state $|\Phi\rangle\langle \Phi|$ on the two boundary systems that are being connected. The result of connecting the systems $B$ and $C$ of two operations  $\{M^{AB}_i\}_{i\in O}$ and $\{N^{CD}_j\}_{j\in Q}$ is a new operation $\{L^{AD}\}_{ij\in O\times Q}$, where
\begin{gather}
 L^{AD}_{ij}=d^Ad^D\frac{\tr_{BC}\left[|\Phi\rangle\langle\Phi|^{BC}(M^{AB}_i\otimes N^{CD}_j)\right]}{\tr\left[|\Phi\rangle\langle\Phi|^{BC}(\overline{M}^{AB}\otimes \overline{N}^{CD})\right]}, \forall i\in O, j\in Q.   \label{connect}
\end{gather}
In the special case when $\tr_{BC}\left[|\Phi\rangle\langle\Phi|^{BC}(\overline{M}^{AB}\otimes \overline{N}^{CD})\right]= 0^{AD}$, the result is defined as the null operation $\{0^{AD}\}$.\\

(3) A network is an arbitrary graph whose vertices are operations and whose edges are wires (e.g., Fig.~\ref{network}). The events in any network have joint probabilities which depend only on the specification of the network. Since a network has no open wires, it amounts to an operation from the trivial system to itself, $\{p_k\}_{k\in O}$ (which can be obtained according to the previous rule). The probabilities for the different outcomes of the network are exactly $p_k$.

Equivalently, using the process operator (or state) of the wires in a network, the joint probabilities for the events in a network (e.g., Fig.~\ref{network}) can be written
\begin{gather}
p( i,j,\cdots| \{M^{\cdots}_i\}_{i\in O},\{N^{\cdots}_j\}_{j\in Q}, \cdots; network ) = \notag\\
\frac{ \tr[( M_i^{\cdots}\otimes N_j^{\cdots}\otimes \cdots)W^{\cdots}_{wires}] }{ \tr[( \overline{M}^{\cdots}\otimes \overline{N}^{\cdots}\otimes \cdots)W^{\cdots}_{wires}] }\label{generalrule},
\end{gather}
where $W^{\cdots}_{wires}$ is the tensor product of the entangled states associated with the wires, and all operators are defined on the respective systems as before (which are not explicitly labeled here), except that now there is no distinction between two types of systems. This expression is defined for ${ \tr[( \overline{M}^{\cdots}\otimes \overline{N}^{\cdots}\otimes \cdots)W^{\cdots}_{wires}] }\neq 0$; otherwise the probabilities are postulated to be zero.
The operation in any region of a network can be interpreted as a measurement that the region performs on the state resulting from the event in the complement of the region.

Similarly to Eq.~\eqref{operationupdate}, upon learning or discarding of information about the events in a region in agreement with the closed-box assumption, the description of the operation in that region is most generally updated as
\begin{gather}\label{OPUP}
{M}'^{A}_j=  d^A \frac{\sum_{i\in O} T(j,i){M}_i^{A}} {\sum_{j\in Q} \sum_{i\in O} T(j,i)\tr({M}_i^{A})}\hspace{0.2cm},
\end{gather}
 where $ T(j,i)\geq 0$, $\forall i\in O$, $\forall j \in Q$, $\sum_{j\in Q} T(j,i)\leq 1$, $\forall i\in O$.

 While we have developed the framework in a discrete form, its formulation in terms of regions and boundary systems suggests a natural route for extension to continuous quantum field theory, where, as outlined earlier, the regions can be identified with regions of space-time. In this case, the boundary of each space-time region would be associated with a (generally infinite-dimensional) Hilbert space, and the content of each region would be described by a positive-semidefinite operator on that space (in infinite dimensions, a different normalization would be necessary, or one may consider working with unnormalized operators). Regions could be connected through parts of their boundaries to form new regions, with the operator in the resulting region obtained by an analogue of the rule \eqref{connect}, where the systems $B$ and $C$ now correspond to the two sides of the boundary area through which the regions are connected, and the entangled state of the connecting `wire' is similarly associated with that area (again, in infinite dimensions, a different representation of states may be necessary). In order to obtain closed networks and thereby probabilities, certain regions with partial boundaries  would need to be considered (e.g., regions that only have boundary on one side, such as standard preparations and measurements). Via purification on a larger region, these could be assumed outsourced to the perimeter of the network, whose interior would be a compact region (which does not have to be simply connected).

The picture just outlined corresponds to the general boundary approach to quantum field theory proposed and developed by Oeckl \cite{Oeckl, Oeckl3, Oeckl2}. Here, we will not discuss how to define an actual field theory in this framework, which could involve various subtleties. We note, however, that our framework agrees with the main probability rule proposed in the general boundary approach of Oeckl, while offering several generalizations. In particular, it incorporates the possibility for reflection with respect to a hypersurface by postulating that the tensor product of the Hilbert spaces on both sides contains a measurable state. It also allows more general than unitary dynamics and non-projective measurements on the boundary. 

 \subsection*{Understanding the causal structure of space-time}

By construction, the networks in our theory are in agreement with observation, but they are defined without any pre-specified time orientation. In the regimes where quantum theory has been tested, however, we have the idea of a background space-time over which quantum physics takes place. How could we account for this phenomenon in the framework? One possibility is to introduce the space-time metric as another quantum filed, assuming that its laws of dynamics are such that it is approximately static in the tested regimes of quantum field theory. Another possibility is that the metric could be understood as arising from properties of the operators describing the dynamics of the rest of the fields. Here, we provide arguments in support of this latter conjecture.  


The suggestion that we could recover a notion of time in this formulation may appear surprising at first, because without the distinction between systems of type $1$ and type $2$, an operation corresponding to, say, a unitary transformation from a given input to a given output system is described in exactly the same way as the standard preparation of a bipartite entangled state. But according to the standard interpretation, in the first case we have a channel that transmits information, whereas in the second we have correlations that do not involve signaling. How do we distinguish between these cases? The answer that we propose is that in order to identify the causal structure of space-time, we have to look at compact regions of space-time and consider the operators on the full Hilbert spaces of their boundaries. Heuristically, the idea is that such an operator describes `pure' dynamics in the region as it is not obtained via measurements on any of the fields inside, and hence the correlations that it induces between different points on its boundary can only represent causal relations mediated through the region. The standard preparation of a state does not correspond to an operator of this kind---it is only an effective operator on a subsystem of the boundary of a compact space-time region, obtained conditionally on a measurement on the rest of the boundary.

\begin{figure}
\vspace{0.5 cm}
\begin{center}
\includegraphics[width=8.5cm]{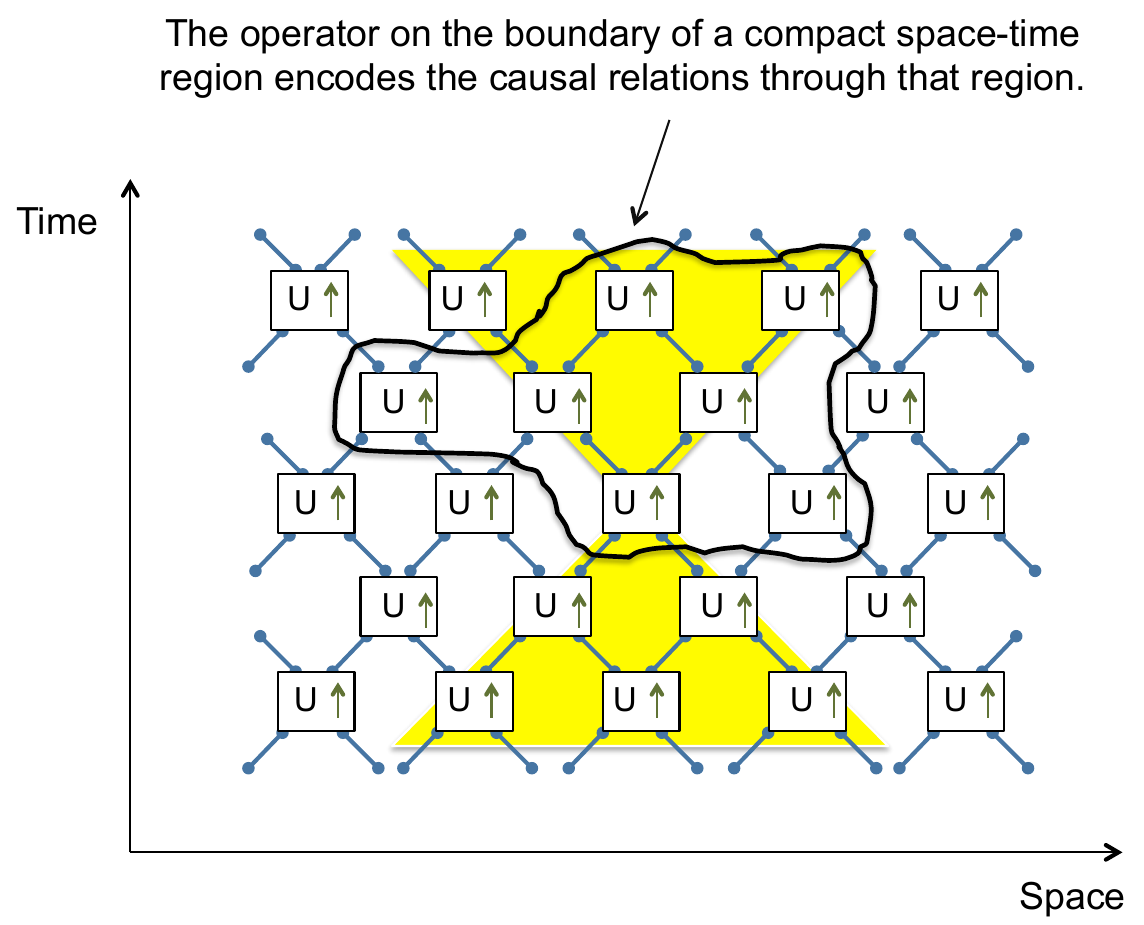}
\end{center}
\vspace{-0.5 cm}
\caption{\textbf{Causal structure from the operators in compact space-time regions.} The causal structure of space-time, defined by the light cones at each point (example in yellow) is expected to be reflected in properties of the operators in compact space-time regions. Here, the operator on the boundary of the region enclosed in black, which corresponds to the circuit inside, would establish correlations between points on the boundary in agreement with the causal structure inside.} \label{toyworld}
\end{figure}

To get intuition about how the causal structure could be manifested in the boundary description, consider a discretized model of a unitary quantum field theory in Minkowski space-time, where the dynamics in a given region of space-time is approximated by a unitary circuit on a lattice, as sketched in Fig.~\ref{toyworld}. While a similar picture may be applicable in more general space-times, we warn that there are subtleties concerning the unitary implementability of the dynamics in quantum field theory in curved space-times (see, e.g., Refs.~\cite{OecklUnitarity, Cortez} and references therein). In this discretized model, the causal structure of space-time is expressed in the structure of the circuit, which imposes limits on the propagation of information in agreement with the `light' cones. One can expect that these limits will be reflected in the correlations that the operator describing the dynamics in a given region of space-time establishes between point on its boundary (Fig.~\ref{toyworld}).

A more direct argument can be given based on considerations about symmetries in the limit of continuous quantum field theory. A natural way to describe dynamics in the boundary approach is through a Feynman integral, which can be applied to arbitrary regions consistently with the rule for composition of regions \cite{Oeckl3}. In the language of our formalism, this means that the components of the operator in a given region in the basis of field configurations on the boundary are formally given in terms of Feynman integrals over the interior. If specifying the operators in all possible regions is equivalent to specifying the Lagrangian density function in the Feynman integral, one could in principle recover the Lagrangian density function from the boundary description. For the tested regimes of quantum field theory, described by the Standard Model, the space-time symmetries of the Lagrangian are the isometries of Minkowski space-time---given by the Poincar\'{e} group---which determine the metric. Therefore, at least in these regimes, it seems that the metric could be understood as arising from symmetries of the dynamics of the other fields, without the need to postulate it additionally. 

 
Since we have no assumption of time in the boundary formulation, one may wonder how a time dimension would appear at all. The time dimension can be thought of as intrinsic in the network framework, corresponding to the fact that the picture over the space-time manifold is a network rather than a state, with states being associated with hypersurfaces. The latter in some sense defines the idea that there is `information flow' through each hypersurface. In this respect, the outlined picture resembles the classical situation where we have a given signature of the metric that defines the existence of a single time dimension, but the concrete time-like directions at each point depend on the concrete metric field.

\section*{Discussion}

We proposed an operational formulation of quantum theory which does not assume a predefined time. The main idea underlying the proposal is an epistemic approach to operational theories, in which an operation represents knowledge about the events in a given region. Combined with a novel isomorphism dependent on time reversal, this approach has allowed us to give a modified formulation of quantum theory that does not refer to time or causal structure, opening up the possibility to treat the causal structure of space-time as dynamical.

Our work provides a new perspective on the informational foundations of quantum theory and the role of causality in it, as well as a general framework within which to study information processing with no causal structure \cite{QGcomputers, Lee, chiribella3, OCB, Chiribella12b, Colnaghi, Araujo, OG, Araujo3, Bennett1, daSilva, GCH, CTC,  Aaronson2,  Baumeler4, Morimae} and explore new potential routes to the unification of quantum theory and general relativity, complementing the field-theoretic approach of Refs.~\cite{Oeckl, Oeckl3, Oeckl2}. We have given a heuristic argument about why we expect that a background causal structure may be possible to infer from the operators describing the dynamics of the other physical degrees of freedom, which calls for a rigorous verification and suggests the search for criteria that link the causal structure in a region to the entanglement properties of the operator on its boundary. In this argument, the physical fields that we are considering are regarded as test fields that do not have effect on the causal structure, but in a gravitational theory the causal structure should itself be dynamical. One way in which this could be enacted in the present framework is by associating gravity with a field similar to the other physical fields. Another possibility is that gravity may be an emergent phenomenon arising from the effective causal structure inferred from the dynamics of other fields, similarly to the way conjectured in the case with a background. In either case, formulating a theory of quantum gravity is expected to incorporate in a suitable way the idea of general covariance \cite{Rovelli}, which is one of the main lessons of general relativity \cite{Gravitation}. The developed operational formulation of quantum theory without predefined time offers a natural framework for exploring this subject.

\begin{acknowledgments}

We thank S. Akibue, P. Arrighi, J. Barrett,  \"{A}. Baumeller, C. B\'{e}ny, C. Branciard, \v{C}. Brukner, N. Brunner, G. Chriribella, B. Coecke, R. Colbeck, F. Costa, G. M. D'Ariano, T. Fritz, C. Giarmatzi, A. Grinbaum, L. Hardy,  I. Ibnouhsein, R. Lal, M. S. Leifer, D. Markham, M. Murao, M. P. M\"{u}ller, K. Ried, P. Perinotti, S. Pironio, M. Pusey, R. W. Spekkens, Z. Wang, and M. Zaopo for discussions and comments. 

This work was supported by the European Commission under the Marie Curie Intra-European Fellowship Programme (PIEF-GA-2010-273119), and by the F.R.S.--FNRS under the Charg\'{e} de recherches (CR) Fellowship Programme and Project T.0199.13.

\end{acknowledgments}

\begin{appendix}



\section{The circuit framework and standard quantum theory}

The basic concept in the circuit framework \cite{HardyCircuit,CDP2} is that of \textit{operation} (also called \textit{test} \cite{CDP2}) with an input and an output system. This is a primitive type of experiment which figuratively can be thought of as performed in an isolated box that by definition can exchange information with other experiments only via the input and output systems---an idea dubbed the `closed-box assumption' \cite{OC2}. An operation with an input system A and an output system B is described by a collection of events $\{\mathcal{M}^{A\rightarrow B}_i\}_{i\in O}$ labeled by an outcome index $i$ taking values in some set $O$. Operations are commonly represented pictorially as boxes with input and output wires (see Fig.~1 in the main text). Operations that have a trivial input system (depicted with no wire) are called \textit{preparations}, and those that have a trivial output system are called \textit{measurements}. The trivial system is denoted by $I$. Operations can be composed in sequence and in parallel to form new operations \cite{Coecke}. An acyclic composition of operations with no open wires, such as the one in Fig.~1, is called circuit. It starts with a set of preparations and ends with a set of measurements, and is equivalent to an operation from the trivial system to itself. By definition, these are the experiments for which we can ascribe well defined outcome probabilities dependent only on the specification of the experiment \cite{HardyCircuit,CDP2}. An OPT in the circuit framework prescribes probabilities for the outcomes of any possible circuit, or equivalently, for any preparation $\{\rho^{I\rightarrow A}_i\}_{i\in O}$ and any measurement $\{E^{A\rightarrow I}_j \}_{j\in Q}$ connected to each other: $p(i,j|\{\rho^{I\rightarrow A}_i\}_{i\in O}, \{E^{A\rightarrow I}_j \}_{j\in Q})\geq 0$, $\sum_{i\in O,j\in Q}p(i,j|\{\rho^{I\rightarrow A}_i\}_{i\in O}, \{E^{A\rightarrow I}_j \}_{j\in Q})=1$.

An OPT is formulated in terms of equivalence classes of operations---if two operations $\{\mathcal{M}^{A\rightarrow B}_i\}_{i\in O}$ and $\{\mathcal{N}^{A\rightarrow B}_i\}_{i\in O}$ give rise to the same joint probabilities when plugged into all possible circuits, they are deemed equivalent. Similarly, if two events $\mathcal{M}^{A\rightarrow B}_i\in\{\mathcal{M}^{A\rightarrow B}_i\}_{i\in O}$ and $\mathcal{N}^{A\rightarrow B}_j\in\{\mathcal{N}^{A\rightarrow B}_j\}_{j\in Q}$ associated with two different operations yield the same joint probabilities with other events in all possible circuits, they are deemed equivalent. The equivalence classes of events are called \textit{transformations} \cite{CDP2}. In the cases of preparation and measurement events, they are called \textit{states} and \textit{effects}, respectively. A theory thus prescribes a joint probability $p(\rho^{I\rightarrow A},E^{A\rightarrow I})$ for every state $\rho^{I\rightarrow A}$ and effect $E^{A\rightarrow I}$, so states can be thought of as real functions on effects and vice versa.

In the case of quantum theory, a system $A$ is associated with a Hilbert space $\mathcal{H}^A$ of dimension $d_A$ (we assume finite-dimensions Hilbert spaces). A composite system $XY$ has the tensor-product Hilbert space $\mathcal{H}^X\otimes \mathcal{H}^Y$, and the trivial system $I$ corresponds to the 1-dimensional Hilbert space $\mathbb{C}^1$. A transformation from $A$ to $B$ is a completely positive (CP) and trace-nonincreasing map $\mathcal{M}^{A\rightarrow B}: \mathcal{L}({\mathcal{H}^A})\rightarrow \mathcal{L}({\mathcal{H}^B})$, where $\mathcal{L}({\mathcal{H}^X})$ denotes the space of linear operators over a Hilbert space $\mathcal{H}^X$.  Every such map can be written in the form \cite{Kraus} $\mathcal{M}^{A\rightarrow B} (\cdot) = \sum_{\alpha=1}^{d_Ad_B} K_{\alpha} (\cdot) K_{\alpha}^{\dagger}$, where $\{K_{\alpha}\}_{\alpha=1}^{d_Ad_B}$ are linear maps, $K_{\alpha}: \mathcal{H}^A\rightarrow \mathcal{H}^B$, called the \textit{Kraus operators}. A quantum operation in the standard formulation of quantum theory is a collection of CP maps $\{\mathcal{M}^{A\rightarrow B}_i\}_{i\in O}$, whose sum $\sum_{i\in O} \mathcal{M}^{A\rightarrow B}_i = \overline{\mathcal{M}}^{A\rightarrow B}$ is a CP and trace-preserving (CPTP) map. States $\rho^{I\rightarrow A}$ are thus CP maps with Kraus operators of the form $\{ |\psi_{\alpha}\rangle^A\}_{\alpha=1}^{d_A}$, where $|\psi_{\alpha}\rangle^A$ are (generally unnormalized) vectors in $ \mathcal{H}^A$, i.e., $\rho^{I\rightarrow A} (\cdot)= \sum_{\alpha=1}^{d_A} |\psi_{\alpha}\rangle (\cdot) \langle \psi_{\alpha}|^A$, where the input $(\cdot)$ stands for a number in $\mathbb{C}^1$. States are therefore isomorphic to positive semidefinite operators $\rho^A\in{\mathcal{L}(\mathcal{H}^A)}$, $\rho^{I\rightarrow A}\leftrightarrow \rho^A= \sum_{\alpha=1}^{d_A} |\psi_{\alpha}\rangle \langle \psi_{\alpha}|^A$, and this is how they are commonly represented. A preparation is then represented by a set of positive semidefinite operators $\{\rho^A_i\}_{i\in O}$ with the property $\sum_{i\in O}\tr(\rho^A_i) = 1$. Effects are similarly described by CP maps with Kraus operators of the form $\{ \langle \phi_{\alpha}|^A\}_{\alpha=1}^{d_A}$, where $\langle \phi_{\alpha}|^A $ are vectors in $ \mathcal{H}^{A^*}$, the dual Hilbert space of  $\mathcal{H}^A$: $E^{A\rightarrow I}(\cdot)= \sum_{\alpha=1}^{d_A} \langle\phi_{\alpha}| (\cdot)|\phi_{\alpha} \rangle^A$, where the input  $(\cdot)$ stands for an operator in $\mathcal{L}(\mathcal{H}^A)$. These are also isomorphic to operators $E^A\in\mathcal{L}(\mathcal{H}^A)$, $E^{I\rightarrow A}\leftrightarrow E^A= \sum_{\alpha=1}^{d_A} |\phi_{\alpha}\rangle \langle \phi_{\alpha}|^A$, and this is how they are represented. The trace-preserving condition implies that a measurement is described by a set of positive semidefinite operators $\{E^A_j\}_{j\in Q}$ that form a positive operator-valued measure (POVM), i.e., $\sum_{j\in Q} E^A_j=\id^A$. In this representation, the joint probability for a pair of state and effect is 
\begin{gather}
p(\rho^{I\rightarrow A}_i, E^{A\rightarrow I}_j) =\tr(\rho^A_i E^A_j). \label{standardrule}
\end{gather}
It is important to stress that states and effects are associated with elements of vector spaces that are dual to each other \cite{CDP2}. The canonical description in terms of vectors in the same space of Hilbert-Schmidt operators is a convenient representation based on the bilinear form \eqref{standardrule}.

\section{Physical admissibility and equivalence of the cyclic circuit framework and the generalized process framework}

Consider a scenario in which a set of separate operations  $\{M^{A_1B_2}_i\}_{i\in O}, \{N^{C_1D_2}_j\}_{j\in Q},\cdots$ have joint probabilities given by 
\begin{gather}
p(i,j, \cdots|\{M^{A_1B_2}_i\}_{i\in O}, \{N^{C_1D_2}_j\}_{j\in Q},\cdots; w)
=\notag\\
\frac{\tr\left[ W^{A_1B_2C_1D_2\cdots}\left(M^{A_1B_2}_i\otimes N^{C_1D_2}_j\otimes\cdots\right)\right] } {\tr\left[ W^{A_1B_2C_1D_2}\left(\overline{M}^{A_1B_2}\otimes \overline{N}^{C_1D_2}\otimes\cdots\right) \right]}, \label{pmformSI}
\end{gather}
\begin{gather}
W^{A_1B_2C_1D_2\cdots}\geq 0, \hspace{0.2cm}\tr W^{A_1B_2C_1D_2\cdots}=1,\label{procondSI}
\end{gather}
where for ${\tr\left[ W^{A_1B_2C_1D_2}\left(\overline{M}^{A_1B_2}\otimes \overline{N}^{C_1D_2}\otimes\cdots\right) \right]}= 0$, the probabilities are defined to be zero. 

A universal way of realizing such a scenario with an arbitrary$W^{A_1B_2C_1D_2\cdots}$ is the following. Prepare a standard density operator (deterministic state) $\rho^{A_1B'_1C_1D'_1\cdots}$, which has identical components to those of $W^{A_1B_2C_1D_2\cdots}$ but is defined over some forward oriented systems $B'_1$, $D'_1$, $\cdots$ which are copies of $B_2$, $D_2$, $\cdots$. The systems $A_1$, $C_1$, $\cdots$ are fed into the inputs of the operations, while the systems $B'_1$, $D'_1$, $\cdots$ together with the corresponding output systems of the operations (let us call them $B_1$, $D_1$, $\cdots$) are subject to a measurement of which an outcome with measurement operator proportional to $S^{B_1}|\Phi^+\rangle\langle \Phi^+|^{B_1B_1'}{S^{B_1}}^{\dagger}\otimes S^{D_1}|\Phi^+\rangle\langle \Phi^+|^{D_1D_1'}{S^{D_1}}^{\dagger} \otimes\cdots$ is post-selected. The latter effectively `teleports' the part of the initial density operator that lives on $B'_1$, $D'_1$, $\cdots$ onto the systems $B_2$, $D_2$, $\cdots$. In other words, the class of joint probabilities between separate operations of the form \eqref{pmformSI} is equivalent to the class of such probabilities obtainable with acyclic circuits with post-selected operations.  In order for this procedure to be geometrically possible, it is sufficient that we have a space-time of dimension 2+1 or higher. This guarantees that all systems that are measured at the final time occupy a space of dimension at least 2 so that any necessary pairwise interactions can be realized without contradiction (e.g., if all systems are ordered in a 1-dimensional chain, they can be accessed as desired from the additional dimension). This can also be seen to follow from the fact that any graph can be embedded in a 3-dimensional space without crossing of its edges. This means that, modulo the transformation that describes time reversal (which according to the present understanding of quantum mechanics can be described by unitary $S$), all correlations \eqref{pmformSI} are physically admissible in the generalized sense that we consider.

The above procedure can also be interpreted as embedding the separate operations in a circuit with a cycle, where the systems $A_1$, $B_2$, $\cdots$, of the separate operations are connected, respectively, to the systems $A'_2$, $B'_1$, $\cdots$, of a single-outcome operation described by the operator
\begin{gather}
R^{A'_2B'_1\cdots}=\\
 d_{A'_2B'_1\cdots} \frac{S^{A'_2}\otimes S^{B'_1} \otimes \cdots (W^T)^{A'_2B'_1\cdots}  {S^{A'_2}}^{\dagger}\otimes {S^{B'_1}}^{\dagger} \otimes \cdots} { \tr(S^{A'_2}\otimes S^{B'_1} \otimes \cdots (W^T)^{A'_2B'_1\cdots}  {S^{A'_2}}^{\dagger}\otimes {S^{B'_1}}^{\dagger} \otimes \cdots )},\notag
\end{gather}
which implements a transformation from the outputs of the separate operations to their inputs. Notice that this is itself a valid operation associated with the region connecting the different operations according to the closed-box assumption, since it can be implemented in an isolated fashion. 
Also, any correlations between operations embedded in some circuit where cycles are allowed are always of the form \eqref{pmformSI}. In other words, the class of correlations \eqref{pmformSI} are equivalent also to the class of correlations obtainable by circuits with cycles.

Let us also show that all cyclic circuits are obtainable in practice with the use of post-selection. A way of creating an arbitrary circuit is the following. Using the teleportation method outlined above, we can realize an operator $W$ equal to the one describing the wires connecting the different operations in the circuit. Strictly speaking, however, this is not the desired circuit because here the connection between the regions of the original operations would be mediated by effective back-in-time identity channels, which are not the same things as wires. Indeed, `genuine' wires are associated with the immediate point of contact of regions. To remedy this, we can simply redefine the regions occupied by the operations in the circuit, for instance by extending each of them along the identity channels connected to its outputs until it reaches the inputs of the operations it is supposed to connect to. In this way, the regions of the operations will be directly connected to each other.

Finally, within the general approach to operations that we are considering, any physical circumstances in which the correlations between a set of separate operations is described by the expression \eqref{pmformSI}, even if they are not assumed to arise from a circuit, can be interpreted as a circuit with a cycle. This is because, if the individual operations satisfy the closed-box assumption, then the collection of events that define the circumstances in which the operations take place, and hence the operator $W$, must be external to the boxes of the operations and so they would define a valid operation in the exterior of their boxes, which can be seen as a box directly connected to the inputs and outputs of the operations, as illustrated in the Section `Existence of cyclic circuits without post-selection'. 

\textit{Note}. It is possible to conceive theories defined in the language of Eq.~\eqref{pmformSI} in which the operator $W$ is not associated with events in any exterior region. For example, we may imagine that the separate operations occupy regions of a space with indefinite topology and thereby have indefinite connections, without there existing any additional region in that space that could contain an operation completing the experiment to a cyclic circuit. In that sense, Eq.~\eqref{pmformSI} can be regarded as a more general starting point than the framework of circuits with cycles. However, one can in principle always extend the underlying space by postulating the existence of a region that completes the experiment to a circuit.



\end{appendix}





\end{document}